# Rainich theory applied to m-rank tensors in n-dimensions.

**Alberto C Balfagón**.

Sección de Física Teórica y Aplicada. Departamento de Ingeniería Industrial, Instituto Químico de Sarriá, Barcelona (Spain)

E-mail: albert.balfagon@iqs.edu

**Abstract.**
We show a tensorial-computational way to find out conditions that must fulfil an m-rank tensor in arbitrary dimension in order to be algebraically the energy-momentum tensor of some field.
We apply in this paper our method to three 2-rank tensors: electromagnetic, mass less scalar field and perfect fluid tensors. Finally we apply the Rainich theory to some examples.

PACS numbers: 04.20.-q, 04.20.Cv, 02.40.-k

Rainich theory put forward the necessary and sufficient algebraic conditions that must fulfil an energy-momentum tensor of some field in order to be equivalent to the energy-momentum tensor of some other specific field. First conditions were established by Rainich [1] giving the conditions of a Maxwell field. Later other authors gave the conditions to other physical fields but limiting themselves, due to the way they take to proof the conditions, to 2-rank energy-momentum tensors in 4 or n dimensions or to 4-rank energy-momentum tensors in 4 dimensions. So Coll and Ferrando [2] studied the perfect fluid, Bergqvist and Senovilla [3] developed a more, and a bit complicate, general way to find the conditions applied principally to 2-rank tensors in n-dimensions, using antisymmetrisation (dimensional properties) Berqvist and Höglund [4] examined the conditions for general p-forms fields in n dimensions, finally Bergqvist and Lankinen [5] using spinorial methods studied the 4-rank Bel-Robinson tensor.

In section 1 we present a general way to find the Rainich theory of any m-rank tensor in arbitrary dimensions using a general tensor polynomial, with initially arbitrary coefficients, and where we will substitute the energy-momentum tensor of the field under study in order to find the conditions of its Rainich Theory. We apply in section 2 the method to some physical fields and finally in section 3 we present some examples using the Rainich theory. We shall use the signature $+,-,-,...,-$ of the metric.

## 1. General polynomial.

Given one arbitrary field $A_{\alpha\beta\gamma...}$ we can construct its m-rank $T_{\alpha\beta\gamma...}$ energy-momentum tensor or its superenergy tensor using the method developed by Senovilla [6]. With this m-rank tensor we can build a linear combination with terms that contains two o more T tensors and the metric $g_{\alpha\beta}$ of the Lorentzian manifold, with p free indexes. The general polynomial must contain all possible combinations of free indexes. In order to find the Rainich conditions of the field we replace in the polynomial the T tensors with its full expression with the A field. This new expression must be simplified using the field algebraic properties and in some cases it is necessary the use of dimensional properties that are obtained through an antisimetrization process. Finally the value of the coefficients in the polynomial can be fixed in order the lineal combination is verified. If there is no solution this means that with the polynomial used does not exist conditions that relate the tensor T, but we can still try to find out relations changing the polynomial: adding more T tensors or making the covariant derivatives of the T tensor in each polynomial term.

We have done all the calculations using a symbolic program developed by Balfagón, Jaén and Castellví [7,8] who proved its performance in [9].

What is studied next is the application of the method over two of the most important cases.

**Rank-2 tensors.**

We analyse the product of at most two 2-rank, symmetric T tensors resulting in 2 free indexes. The general polynomial would be:

$$A[1]T_{\alpha\gamma}T_{\beta}^{\;\gamma} + A[2]T_{\gamma\delta}T^{\gamma\delta}g_{\alpha\beta} + A[3]T_{\gamma}^{\;\gamma}T_{\delta}^{\;\delta}g_{\alpha\beta} + A[4]T_{\alpha\beta}T_{\gamma}^{\;\gamma} + \\ A[5]T_{\alpha\beta} + A[6]T_{\gamma}^{\;\gamma}g_{\alpha\beta} + A[7]g_{\alpha\beta} = 0 \quad (1)$$

Where the A[i] coefficients would have dimensions to give physical sense to polynomial (1). So, for example, A[7] coefficient will have the necessary dimensions in order it can be added to the other polynomial terms that only depend on T.

A tensor T that fulfil (1) must have a trace given by:

$$T_{\gamma}^{\;\gamma} = 2(n + A[4])^{-1} \bullet (-(A[5] + nA[6]) \pm \\ [(A[5] + nA[6])^2 - 4(nA[3] + A[4])((A[1] + nA[2])T_{\gamma\delta}T^{\gamma\delta} - nA[7])]^{1/2}) \quad (2)$$

Where n is the manifold dimension.

**Rank-4 tensors.**

In the case of a totally symmetric, trace free, 4-rank T tensor, like the Bel-Robinson tensor, we must first of all choose the number of free indexes we want. If we analyse the product of at most two tensors we have three possibilities:

   a) Two free indexes. The polynomial would be:

$$A[1]T_{\alpha\gamma\delta\mu}T_{\beta}^{\;\gamma\delta\mu} + A[2]T_{\gamma\delta\mu\eta}T^{\gamma\delta\mu\eta}g_{\alpha\beta} = 0 \quad (3)$$

Taking the trace of (3) we find that A[1] must be equal to $nA[2]$, so (3) would be:

$$T_{\alpha\gamma\delta\mu}T_{\beta}{}^{\gamma\delta\mu} = \left(1/n\right)T_{\gamma\delta\mu\eta}T^{\gamma\delta\mu\eta}g_{\alpha\beta} \qquad (4)$$

This is a relation fulfilled by the Bel-Robinson tensor in four (n = 4) dimensions [10], it is a necessary, but not sufficient condition, that must accomplish a tensor T to be algebraically equivalent to the Bel-Robinson tensor. This is so because we have found condition (4) without any physical information of tensor A neither his energy-momentum T, condition (4) is an algebraic property fulfilled by all totally symmetric, trace free, 4-rank T tensor. If we want find out the necessary and sufficient condition (if it exist, that is the situation) we must start by working with more than two free indexes.

b) Four free indexes. The polynomial terms are grouped in three kind of products:

$$T_{\alpha\beta\mu\eta}T_{\gamma\delta}{}^{\mu\eta}; \quad T_{\gamma\mu\eta\rho}T_{\delta}{}^{\mu\eta\rho}g_{\alpha\beta}; \quad T_{\mu\eta\rho\zeta}T^{\mu\eta\rho\zeta}g_{\alpha\beta}g_{\gamma\delta}$$

c) Six free indexes. The polynomial terms are grouped in four different groups:

$$T_{\alpha\beta\gamma\omega}T_{\delta\mu\eta}{}^{\omega}; \quad T_{\alpha\beta\omega\zeta}T_{\gamma\delta}{}^{\omega\zeta}g_{\mu\eta};$$
$$T_{\alpha\omega\zeta\psi}T_{\beta}{}^{\omega\zeta\psi}g_{\gamma\delta}g_{\mu\eta}; \quad T_{\omega\xi\psi\zeta}T^{\omega\xi\psi\zeta}g_{\alpha\beta}g_{\gamma\delta}g_{\mu\eta} \qquad (5)$$

Bergqvist and Lankinen [5] found out that, in four dimensions, the necessary and sufficient condition for a tensor T to be algebraically equivalent to Bel-Robinson tensor was obtained by terms of the sort of (5).

## 2. Examples of application to rank-2 tensors.

In this section we show some applications of the method to 2-rank, symmetric T tensors. We need for each case the energy-momentum tensor expressed as a function of the physical field at n dimension. Where it were needed dimensional properties will be used.

**Electromagnetic tensor.**

Electromagnetic field is given by the Maxwell tensor $F_{\alpha\beta}$ a 2-form. The energy-momentum tensor at n-dimensions is:

$$T_{\alpha\beta} = -F_{\alpha\gamma}F_{\beta}{}^{\gamma} + \frac{1}{4}g_{\alpha\beta}F_{\gamma\delta}F^{\gamma\delta} \qquad (6)$$

The trace being given by:

$$T_{\gamma}{}^{\gamma} = \frac{n-4}{4}F_{\lambda\delta}F^{\lambda\delta} \qquad (7)$$

We now have to substitute expression (6) in (1) but before doing this we should realize that, in this case, polynomial (1) can be divided in two independent expressions, the first one will contain terms with the product of two T tensors that will product terms with 4 Maxwell tensors. The second one will have terms with at most one T tensor that will product terms with 2 or any Maxwell tensors:

$$A[1]T_{\alpha\gamma}T_{\beta}{}^{\gamma} + A[2]T_{\gamma\delta}T^{\gamma\delta}g_{\alpha\beta} + A[3]T_{\gamma}{}^{\gamma}T_{\delta}{}^{\delta}g_{\alpha\beta} + A[4]T_{\alpha\beta}T_{\gamma}{}^{\gamma} = 0 \qquad (8)$$

$$A[5]T_{\alpha\beta} + A[6]T_\gamma{}^\gamma g_{\alpha\beta} + A[7]g_{\alpha\beta} = 0 \qquad (9)$$

Depending on dimension we get the following results:

*Theorem 1.* In 4 dimensions, a tensor $T_{\alpha\beta}$ must fulfil conditions (10) in order to be algebraically the energy-momentum tensor of a Maxwell field.

$$T_\gamma{}^\gamma = 0 \quad \text{and} \quad T_{\alpha\gamma}T_\beta{}^\gamma = \frac{1}{4}g_{\alpha\beta}T_{\gamma\delta}T^{\gamma\delta} \qquad (10)$$

Proof: replacing (6) in (8) and using (7) we obtain an expression that needs the use of a dimensional property to find the non nulls A[i] coefficients. We give the dimensional property used because of the importance of this case:

$$F_{\gamma\delta}F^{\gamma\delta}F_{\mu\eta}F^{\mu\eta}g_{\alpha\beta} = 8F_{\gamma\delta}F^\gamma{}_\beta F^\delta{}_\mu F^\mu{}_\alpha + 4F_{\gamma\delta}F^{\gamma\delta}F_{\mu\alpha}F^\mu{}_\beta + 2F_{\gamma\delta}F^\gamma{}_\mu F^\delta{}_\eta F^{\mu\eta}g_{\alpha\beta}$$

It is necessary a dimensional property to get the theorem, but if we decompose the Maxwell field in its electric and magnetic part then we do not need any dimensional property. We won't use this decomposition in the rest of results.

Replacing (6) in (9) does not yield anything.

*Theorem 2.* In 3 dimensions, a tensor $T_{\alpha\beta}$ must fulfil conditions (11) in order to be algebraically the energy-momentum tensor of a Maxwell field.

$$T_\gamma{}^\gamma = \pm\left(\frac{1}{3}T_{\alpha\beta}T^{\alpha\beta}\right)^{1/2} \quad \text{and} \quad T_{\alpha\gamma}T_\beta{}^\gamma = \frac{1}{3}g_{\alpha\beta}T_{\gamma\delta}T^{\gamma\delta} \qquad (11)$$

Proof: replacing (6) in (8) and with dimensional properties we get the A[i] coefficients then using (2) with A[5]=A[6]=A[7]=0 we get the theorem conditions. Substituting (6) in (9) does not yield anything.

*Theorem 3.* In 2 dimensions, a tensor $T_{\alpha\beta}$ must fulfil one of the two equivalent conditions (12) in order to be algebraically the energy-momentum tensor of a Maxwell field.

1) $T_\gamma{}^\gamma = \left(2T_{\alpha\beta}T^{\alpha\beta}\right)^{1/2} \quad$ and $\quad T_{\alpha\gamma}T_\beta{}^\gamma = \frac{1}{2}g_{\alpha\beta}T_{\gamma\delta}T^{\gamma\delta}$

2) $T_{\alpha\beta} = \frac{1}{2}g_{\alpha\beta}T_\gamma{}^\gamma$

(12)

Proof: the first set of conditions is obtained replacing (6) in (8) and with dimensional properties we get the A[i] coefficients then using (2) with A[5]=A[6]=A[7]=0 we find the theorem conditions.

Substituting (6) in (9) yield the second set of conditions. It is straightforward to pass from this condition to the first one, so the two sets of conditions are equivalents.

*Theorem 4.* In dimensions greater than 4, does not exist conditions using polynomial (1) that let a tensor $T_{\alpha\beta}$ be algebraically the energy-momentum tensor of a Maxwell field.

Proof: in dimension greater than four does not exist any dimensional property that could be applied to (1) in order to get non nulls A[i] coefficients.

**Mass less scalar field.**
The mass less scalar field is given by the (covariant-) derivative of a scalar function $\phi$, its minimally coupled energy-momentum tensor in n dimensions is [3,11]:

$$T_{\alpha\beta} = \phi_{;\alpha}\phi_{;\beta} - \frac{1}{2}g_{\alpha\beta}\phi_{;\gamma}\phi^{;\gamma} \qquad (13)$$

Its trace is given by:

$$T_\gamma^{\ \gamma} = \left(\frac{2-n}{2}\right)\phi_{;\delta}\phi^{;\delta} \qquad (14)$$

As in the electromagnetic field, the polynomial (1) can be divided in two polynomials (8) and (9). Replacing (13) in (9) does not give any solution.
Taking (2) with A[5]=A[6]=A[7]=0 and using (13) and (14) give us a general result that if (1) must have a solution then it is necessary the following trace condition:

$$T_\gamma^{\ \gamma} = \pm(n-2)\left(\frac{T_{\alpha\beta}T^{\alpha\beta}}{n}\right)^{1/2} \qquad (15)$$

*Theorem 5.* In n dimensions a tensor $T_{\alpha\beta}$ must fulfil conditions (16) in order to be algebraically the energy-momentum tensor of a minimally coupled mass less scalar field $\phi$.

$$T_\gamma^{\ \gamma} = \pm(n-2)\left(\frac{T_{\alpha\beta}T^{\alpha\beta}}{n}\right)^{1/2} \quad \text{and} \quad T_{\alpha\gamma}T_\beta^{\ \gamma} = \frac{1}{n}T_{\mu\eta}T^{\mu\eta}g_{\alpha\beta} \qquad (16)$$

Proof: replacing (13) and (14) in (8) and using (15) it is a straightforward calculation to get the theorem conditions without using any dimensional property.

**Perfect fluid field.**
The perfect fluid energy-momentum tensor is given by:
$$T_{\alpha\beta} = (\rho + p)U_\alpha U_\beta - pg_{\alpha\beta} \qquad (17)$$

Where $\rho$ and $p$ are the density and pressure of the fluid and U is a temporal vector: $U_\alpha U^\alpha = 1$.

Before substituting (17) in (1) we can see that with this energy-momentum tensor, some terms in (1) will be proportional, therefore (1) can be simplified before starting the calculus. Analysing all the terms in (1) we realize that polynomial (1) can be reduced to, redefining the terms used in (1):

$$A[1]T_{\alpha\gamma}T_\beta^{\ \gamma} + A[2]T_{\alpha\beta} + A[3]g_{\alpha\beta} = 0 \qquad (18)$$

This relation between the terms in (1) can be stated as:

$$nT_{\alpha\beta}T^{\alpha\beta} - \left(T_\alpha^{\ \alpha}\right)^2 \geq 0$$

Where n is the dimension of the manifold.

After replacing (17) in (18) we find out the following result [3]:

*Theorem 6.* In n dimensions a tensor $T_{\alpha\beta}$ must fulfil conditions (19) in order to be algebraically the energy-momentum tensor of a perfect fluid.

$$T_{\alpha\gamma}T_{\beta}^{\ \gamma} = \lambda T_{\alpha\beta} + \delta g_{\alpha\beta} \quad \text{and} \quad nT_{\alpha\beta}T^{\alpha\beta} - \left(T_{\alpha}^{\ \alpha}\right)^2 \geq 0 \qquad (19)$$

Where $\lambda = (\rho - p)$ and $\delta = \rho \cdot p$. In other words, $\lambda$ and $\delta$ would be a combination of the density and pressure of the field that is now taken as a perfect fluid.

Proof: it is a very simple calculation after substituting (17) in (18) to get the theorem conditions without using any dimensional property.

## 3. Rainich theory applied to some fields.

In this section we shall apply the Rainich theory to the fields that have been studied before. Our only intention is to show one way this theory can be applied, without intending to obtain any physical consequence.

The idea is to find out the conditions that a field must fulfil in order its energy-momentum tensor could be considered as, for example, a mass less scalar field and simultaneously a perfect fluid one. Therefore we can study three relations.

**Mass less scalar field and perfect fluid.**
Starting with the perfect fluid, we have to replace (17) in the conditions (16). After a straightforward calculation we obtain that a perfect fluid will have an energy-momentum tensor algebraically equivalent to an energy-momentum tensor of a mass less scalar field if and only if $\rho = p$.

Replacing the scalar field energy-momentum (13) in the conditions (19) we find that an energy-momentum of a mass less scalar field would be algebraically the energy-momentum tensor of a perfect fluid if and only if we consider that $\phi_{:\alpha}\phi^{:\alpha} = \pm 2\rho$, where $\rho$ would be the density of the perfect fluid and $\rho = p$.

Summing up, the perfect fluid and mass less scalar field energy-momentum tensors can be considered simultaneously algebraically equivalents if and only if we take $\rho = p$ and $\phi_{:\alpha}\phi^{:\alpha} = \pm 2\rho$.

**Electromagnetic field and perfect fluid.**
Doing as before we can demonstrate easily that only in 3 dimensions we can find fields whose energy-momentum tensor could be considered simultaneously algebraically equivalent to the perfect fluid and electromagnetic energy-momentum tensor.

In 3 dimensions, after replacing (17) in (11) we find out that $\rho$ must be equal to $p$ in order to fulfil conditions (11).

Replacing (6) in (19), and taking in to account the last condition ($\rho = p$) we find that to fulfil conditions (19) the electromagnetic field must verify:

$$F_{\alpha\beta}F^{\alpha\beta} = \pm 4\rho \quad \text{and} \quad F_{\alpha\gamma}F_{\delta}^{\ \gamma}F_{\beta\lambda}F^{\delta\lambda} = \pm 2\rho F_{\alpha\gamma}F_{\beta}^{\ \gamma} \qquad (20)$$

As a result, in 3 dimensions the perfect fluid and electromagnetic field energy-momentum tensors can be considered simultaneously algebraically equivalents if and only if we take $\rho = p$ and the conditions (20) are verified.

**Mass less scalar field and electromagnetic field.**

In this case the conditions for the mass less scalar field (16) and for the electromagnetic field (10,11,12) are very similar, the only difference if it exist is the trace condition.

Using the trace condition that appears in (16) and in (10,11,12) depending of the manifold dimension, we obtain that the electromagnetic and mass less scalar field energy-momentum tensors can be considered simultaneously algebraically equivalents if and only if we take:

a) For n = 4: $\phi_{:\alpha}\phi^{:\alpha} = 0$ and $F_{\alpha\beta}F^{\alpha}{}_{\delta}F^{\beta}{}_{\gamma}F^{\delta\gamma} = \frac{1}{4}\left(F_{\alpha\beta}F^{\alpha\beta}\right)^2$.

b) For n = 3: $\phi_{:\alpha}\phi^{:\alpha} = \pm\frac{1}{2}F_{\alpha\beta}F^{\alpha\beta}$ and $F_{\alpha\beta}F^{\alpha}{}_{\delta}F^{\beta}{}_{\gamma}F^{\delta\gamma} = \frac{1}{2}\left(F_{\alpha\beta}F^{\alpha\beta}\right)^2$

c) For n = 2: $\phi_{:\alpha}\phi^{:\alpha} = F_{\alpha\beta}F^{\alpha\beta} = 0$

## 4. Conclusions.

We have shown a tensorial way to find out a set of necessary conditions that must fulfil an m-rank tensor in n dimensions to be algebraically the energy-momentum tensor of other field.

We also have found the conditions in arbitrary dimensions for three 2-rank energy-momentum tensors: electromagnetic, mass less scalar field and perfect fluid tensors.

There is an important application of the method, it is the case of the Bel-Robinson tensor. Bergqvist and Lankinen [5] found the conditions in four dimension using spinors. With the method presented in this paper it would be possible to show the conditions in a tensorial way and find the conditions for arbitrary dimension, it is only a matter of computational power. In this moment I am unable to perform the computational calculus with the hardware I am using, it is for this reason that I have no presented any proof of [5].

Finally we have shown a simple examples of application of the Rainich theory.